\pdfoutput=1
\documentclass[aps,pre,preprint,superscriptaddress]{revtex4-1}
\usepackage{graphicx}
\usepackage[utf8x]{inputenc}

\begin{document}

\title{Modelling proton transfer in water molecule chains}

\author{A. Korzhimanov} 
\affiliation{Department of Physics, Ume{\aa} University, SE-901 87 Ume{\aa}, Sweden}
\altaffiliation[Also at: ]{Institute of Applied Physics, Russian Academy of Sciences, 603950 Nizhny Novgorod, Russia}

\author{M. Marklund}
\affiliation{Department of Physics, Ume{\aa} University, SE-901 87 Ume{\aa}, Sweden}

\author{T. Shutova} 
\affiliation{Department of Plant Physiology, Ume{\aa} Plant Science Centre, Ume{\aa} University, SE-901 87 Ume{\aa}, Sweden}

\author{G. Samuelsson}
\affiliation{Department of Plant Physiology, Ume{\aa} Plant Science Centre, Ume{\aa} University, SE-901 87 Ume{\aa}, Sweden}

\date{\today}

\begin{abstract}
The process of protons transport in molecular water chains is of fundamental interest for many biological systems. Although many features of such systems can be analyzed using large-scale computational modeling, other features are better understood in terms of simplified model problems. Here we have tested, analytically and numerically, a model describing the classical proton hopping process in molecular water chains. In order to capture the main features of the proton hopping process in such molecular chains, we use a simplified model for our analysis. In particular, our discrete model describes a 1D chain of water molecules situated in an external protein channel structure, and each water molecule is allowed to oscillate around its equilibrium point in this system, while the protons are allowed to move along the line of neighboring oxygen atoms. The occurrence and properties of nonlinear solitary transport structures, allowing for much faster proton transport, are discussed, and the possible implications of these findings for biological systems are emphasized.
\end{abstract}

\pacs{87.10.-e}

\maketitle

\section{Introduction}
In many biological reactions, for example in trans-membrane electron transport in proteins, or protein complexes \cite{Stuchebrukhov.jtcc.2003, Cramer.Knaff}, like cytochrome oxidase \cite{Wikstrom.cosp.1998}, bc1 complex \cite{Berry.arb.2000}, or photosynthetic reaction centers (PRC) \cite{Okamura.bba.2000}, proton transport is an important part of the overall reaction. In general, an understanding of charge transport on the micro- and nanoscale is of great interest in a wide variety of fields (see, e.g., \cite{Hanggi}). 
The transfer of protons often takes place in special proton conducting channels inside the proteins. These channels consist of several amino acids with protonable groups, facing into the channels, that are able to transfer proton between each other. These groups are connected by a chain of around two to five mobile water molecules; this is the chain along which the proton-transport occurs by the Grotthuss mechanism.
Recently, a putative proton-conducting channel in PS II was identified from structural analyses \cite{Guskov.nsmb.2009, Gabdulkhakov.str.2009}. It was argued that this channel is filled with water and that the proton transport is based on the Grotthuss mechanism. 

It was further proposed that some amino acid residues on the surface of the channel, located in a hydrophilic pocket, may facilitate an efficient proton removal from the water oxidation center, by creating a proton accepting network \cite{Shutova.bba.2007}. Whatever organization the proton conducting pathway may have, it is believed that the removal from the catalytic Mn$_4$Ca metal center of the protons formed during the oxidation of water is highly important, in order to optimize the energetics for electron abstraction from the Mn$_4$Ca catalytic center \cite{XX, YY}. Even small delays or disturbances in the photochemical reactions will ultimately lead to photoinduced destruction of the reaction centers. 

Unfortunately, based on the known structure of the channel it is hard to make precise predictions how proton transfer actually occurs, and thus different mechanisms could be considered.

Proton transfer in water has been extensively studied numerically \cite{Marx.cphc.2006}. It has been shown that the transport of protons in chains of water could occur either as a random walk between water molecules or as propagation of solitary structures \cite{Stuchebrukhov.jtcc.2003, Davydov.jtb.1973, Davydov.jtb.1977, Antonchenko.pssb.1983, Bountis}. In the former case the proton is localized to a single water molecule or to a single hydrogen bond between two molecules. In the latter case the charge associated with the proton is delocalized between several molecules. Thus the process of proton soliton propagation can be treated as an activated single-step process with high barrier of activation, whereas the random walk transfer can be considered as multi-step diffusion with a comparatively low barrier of activation (for each single proton hopping event). The actual rate and activation barrier of the corresponding processes, and therefore which of the processes that actually occur, is strongly dependent on the actual parameters of the system.

In the current work we analyze how some of the most important parameters of the system influence the proton transport. In particular, we investigate the parameter regions that determine the distinction between the above proton transport scenarios. The analysis is based on the physically motivated model that reflects all the basic properties of the transfer process. Such a model can be modified in a straightforward manner in order to encompass further parameters of the process as well to make predictions for more more complicated setups. Finally, we discuss the results in light of certain biological systems.

\section{Basic equations}
We consider a chain of water molecules squeezed into a channel inside a protein. Each molecule is bounded to the protein by means of a hydrogen bond, and the molecule can thus only oscillate weakly around an equilibrium point. For the sake of simplicity we assume that it has only one degree of freedom -- it is only allowed to oscillate along one direction. There are also hydrogen bonds between the neighboring molecules, and the hydrogens are localized between the oxygen atoms. Furthermore, we assume that the proton can move only along the line connecting neighboring oxygens.

All atoms are considered to be described classically so the precise position of each atom can be introduced (for a recent discussion of the classical-quantum efficiency in transport for electrons, see Ref.\ \cite{Briggs-Eisfeld}). At the same time, the forces acting on the atoms should be derived from treating the electrons quantum mechanically. In this work these forces are described by a simple model that, while neglecting detailed structures, takes into account the most significant features of the interaction.

The oxygens atoms are supposed to experience only weak (i.e.\ linear) oscillations. Thus, a harmonic approximation for this motion can be made. As for the hydrogens, DFT calculations shows that a hydrogen between two oxygens in the $\mathrm{H_5O_2^+}$ complex in the most cases can be described as a classical particle moving in the double-well potential. The shape of this potential depends on the distance $d$ between the oxygens. It is also known that this potential transforms into a single-well potential if the distance is small enough (this distance have been shown to have the value $\approx$\,2.5--2.6\,\AA) \cite{lill.jcp.2001}. Moreover, it is known that in hydronium ($\rm H_3O^+$) there are additional interactions between the hydrogens that tend to push them apart. We take all these effects into account using a simple model describing a one-dimensional water chain. This approach is similar to that used by Stuchebrukhov \cite{Stuchebrukhov.PRE.2009}.

Following the above description, we use a Hamiltonian of the form
\begin{equation}\label{hamiltonian}
H(q, Q, \phi, \Phi) = H_O(\phi,\Phi) + H_H(q, Q) + H_{int}(q, \phi)
\end{equation}
here $q$ and $Q$ are the coordinates and momenta of the hydrogens, respectively, $\phi$ and $\Phi$ are the coordinates and momenta of the oxygens, respectively, $H_O, H_H, H_{int}$ are the Hamiltonians describing the motion of the oxygens, hydrogens and their interaction, respectively. We have the following form of these Hamiltonians
\begin{eqnarray}
H_O(\phi,\Phi) & = & \sum\limits_{i=1}^{N}\left[\frac{\Phi_i^2}{2M} +
\frac{M\omega_O^2\left(\phi_i-\phi_i^0\right)^2}{2}\right] \\
H_H(q, Q) & = & \sum\limits_{i=1}^{N+1}\frac{Q_i^2}{2m} +
\sum\limits_{i=1}^{N}\left[\frac{m\omega_H^2\left(q_{i+1} - q_i -
d\right)^2}{2}\right] \\
H_{int}(q, \phi) & = & V_1\left(q_1, \phi_1\right) +
\sum\limits_{i=2}^{N}V\left(q_i, \phi_{i-1}, \phi_i\right) +
V_{N+1}\left(q_{N+1}, \phi_N\right) ,
\end{eqnarray}
where $N$ is the number of molecules involved in the system, $m, M$ are the masses of oxygen and hydrogen, respectively, $k_O = M\omega_O^2$ is the coefficient defining the strength of the oxygen-protein bond, $k_H = m\omega_H^2$ is the coefficient defining the strength of the interaction between neighboring hydrogens, $\phi_i^0$ is the equilibrium position of oxygen $i$ in the absence of all other atoms, and $d$ is the equilibrium distance between the oxygens. The function $V$ is the only non-linear term in this hamiltonian and it describes the motion of the proton in the field of electrons and nuclei of two neighboring molecules. It should be noted that there is one proton more than the number of oxygens, which means that we introduce one hydronium molecule in the chain and the proton transfer will be with respect to the charge associated with that additional proton. The functions $V_1$ and $V_{N+1}$ introduced above define the boundary conditions of our system, and they have the form
\begin{eqnarray}
V_1\left(q_1, \phi_1\right) & = & m\omega_1^2\left(q_1-(\phi_1-d_H)^2\right)^2\\
V_{N+1}\left(q_{N+1}, \phi_N\right) & = & m\omega_1^2\left(q_{N_+1}-(\phi_N-d_H)^2\right)^2
\end{eqnarray}
Here $d_H$ is the equilibrium length of the $\rm O-H$ bond and $\omega_1$ is the frequency of the proton oscillations in the well.
The form of the function $V$ is of particular interest. This should be as simple as possible but still reflect the main features of the interaction: it should be a double-well function for large enough distances between the oxygens, and a single-well function for small distances \cite{lill.jcp.2001}. The distance between each well and the closest oxygen should remain constant for large enough distances, and the second derivative of the function in the minima should be a constant. It can be shown that all these requirements are satisfied by the function
\begin{eqnarray}
V\left(q_i, \phi_i, \phi_{i-1}\right) & = &
\frac{m\omega^2 }{4B\sqrt{(\phi_i - \phi_{i-1} - d_0)^2 + b^2}}
\left(q_i-\frac{\phi_i+\phi_{i-1}}{2}\right)^2
\nonumber\\
&& \times\left[\left(q_i-\frac{\phi_i+\phi_{i-1}}{2}\right)^2 -
B\left(\phi_i - \phi_{i-1} - d_0\right)\right]
\end{eqnarray}
Here $d_0$ is the distance between oxygens for which the double-well potential transforms into a single-well potential, and $b$ is a small parameter that ensures a smooth transition from a double- to a single-well potential. The $\omega$ determines the frequency of the small oscillations in the wells and the coefficient $B$ determines the potential barrier between them. The potential barrier is given by
\begin{equation}\label{Delta_V}
\Delta V = \frac{m\omega^2 B\left(\phi_i - \phi_{i-1} -
d_0\right)^2}{16\sqrt{\left(\phi_i - \phi_{i-1} - d_0\right)^2 + b^2}} ,
\end{equation}
and the frequency of the small oscillations in the well is
\begin{equation}\label{omega}
\omega_V = \omega\sqrt{\frac{\phi_i - \phi_{i-1} - d_0}{\sqrt{\left(\phi_i - \phi_{i-1} - d_0\right)^2 + b^2}}} .
\end{equation}
The distance between minima of the potential grows with the distance between oxygens and is determined by the expression
\begin{equation}\label{a}
a = q_{min2} - q_{min1} = \sqrt{2B\left(d-d_0\right)} .
\end{equation}

\section{Analysis}
As stated above, there are two qualitatively different proton transfer scenarios that could be realized in our model system: 1) random-walk hopping of protons, or 2) ballistic propagation of delocalized soliton. Beside these two scenarios there are also two trivial regimes: t1) a stable setup in which no transport occurs, or t2) free-hopping (in which protons are freely hopping between the neighboring molecules).
In the stable regime (t1) protons don't have enough energy to overcome the potential barrier $\Delta V$ and all atoms are oscillating near their initial positions. There are no proton hops and thus no transport. This regime is observed if the distance $d$ between neighboring oxygens is large. To be more precise, when the following expression
\begin{equation}\label{stable condition}
T + \frac{k_H B(d-d_0)}{2} \ll \Delta V
\end{equation}
is satisfied, we have the stable regime.

Here $T$ is the temperature of the protons. We note that $d$ is the subject to a change in time due to the oxygen oscillations, but it is assumed that those oscillations are weak. The inequality (\ref{stable condition}) can be violated either by increasing the temperature or by increasing the coupling between the protons. Below, we consider both cases independently.

First, let us assume that the temperature dominates over the second term on the left-hand side of the inequality (\ref{stable condition}). As the distance $d$ is decreased we reach a point when the proton energy becomes comparable to the potential barrier. As soon as this happens a proton has the possibility to hop to the neighboring state. If the distance between the water molecules is not small enough these hops will be a random process, and thus the transport would be a random-walk diffusion along the water chain. The rate of diffusion depends on the distance $d$, temperature $T$ and coupling constants $k_O$ and $k_H$.

Let us derive an estimation for diffusion coefficient. It is well known that for a one-dimensional random-walk process, the diffusion coefficient can be written down as
\begin{equation}\label{1d_diff_coeff}
D = \frac{d^2}{2\tau} .
\end{equation}
Here $d$ is the size of one step (in our system it is the distance between two neighboring oxygens) and $\tau$ is the average time between two consecutive steps. The latter can be estimated according to
\begin{equation}\label{hopping_time}
\tau = \omega_V^{-1}\exp\frac{\Delta U}{T} ,
\end{equation}
where $\omega_V$ is the frequency of the hydrogen oscillations in the minima of the function $V$, as given by expression (\ref{omega}), $\Delta U$ is the potential barrier which proton should overcome in order to make a hop, and $T$ is the temperature. The expression (\ref{hopping_time}) follows from the fact that the probability of the proton to have energy greater than $\Delta U$ is $\exp\left(-\frac{\Delta U}{T}\right)$. The main problem here is to estimate $\Delta U$. At zero approximation it just equals to the potential barrier of the $V$ function given by the expression (\ref{Delta_V}). We will also take into account that it is affected by proton interactions so that
\begin{equation}
\Delta U = \Delta V + \Delta U_H
\end{equation}
Considering a single proton in the potential of the neighboring protons (the latter assumed to be fixed), it will have minimal potential energy at the equidistant point between the other two protons. The difference between the minimal proton energy and the minimum of the potential $V$ is given by
\begin{equation}
\Delta U_H = - k_H\left(d^2 - \frac{a^2}{2}\right)
\end{equation}
here $a$ is the same as in (\ref{a}). Additionally, due to the oxygen motion the \emph{effective} distance between neighboring oxygens may be smaller than $d$. The amplitude of their oscillations depends on the temperature and the coupling constant $k_O$ as
\begin{equation}
A_O = \sqrt{\frac{2T}{k_O}}.
\end{equation}
Thus, the effective distance can be estimated to be
\begin{equation}
d_{\rm eff} = d - \sqrt{\langle(\phi_i - \phi_i^0)^2\rangle} = d - \frac{A_O}{\sqrt{2}} = d - \sqrt{\frac{T}{k_O}},
\end{equation}
here angular brackets denotes the ensemble average. A change in the effective distance will affect the potential $V$ and therefore decrease $\Delta V$. Finally, the diffusion coefficient can be given as 
\begin{equation}\label{diff_coeff}
D = \frac{\omega d^2}{2}\exp\left\{\frac{(m\omega^2 B/16)\left(d - \sqrt{\frac{T}{k_O}} - d_0\right) - k_H\left(d^2 - \frac{a^2}{2}\right)}{T}\right\} 
\end{equation}
(here $b$ is neglected as it is assumed to be much smaller than $d - d_0$).

Let us now consider the case (t2). If the distance $d$ is decreased further, such as the proton energy becomes much greater than the potential barrier, the protons begin to freely hop between the molecules as if there were no barriers present in the system. In this case the behavior of the protons in the system becomes considerably more complex as compared to the aforementioned case.

The qualitative description given above can be illustrated by means of numerical simulation, starting from the hamiltonian (\ref{hamiltonian}) in order to obtain the equations of motion. We use a conventional Runge-Kutta method of the fourth order to solve the equations of motion. The number of molecules in the chain was chosen to be $N = 11$, with the hydronium initially situated at the center of the chain. The oxygens were put at the positions $\phi_i^0$ and the hydrogens were situated at the minima of the potential function $V$. In order to introduce a finite temperature into the system, all the atoms in the simulations were given an initial random momentum with a mean energy corresponding to the temperature of interest. In order to have representative results, a Monte-Carlo scheme was applied: several runs (of the order of ten) with the same parameters but different initial conditions were performed.

The parameters were subject to change but for the sake of simplicity we fixed several of them: $\omega^2 = 0.8 \, \mathrm{fs}^{-2}$, $\omega_1^2 = 0.4 \, \mathrm{fs}^{-2}$, $B = 0.2 \, \mathrm{\AA}$, $b = 10^{-6} \, \mathrm{\AA}$, $d_0 = 2.5 \, \mathrm{\AA}$. The temperature determined by initial momenta of the particles was equal to 300~K in all runs.

\begin{figure}
 \includegraphics[width=\columnwidth]{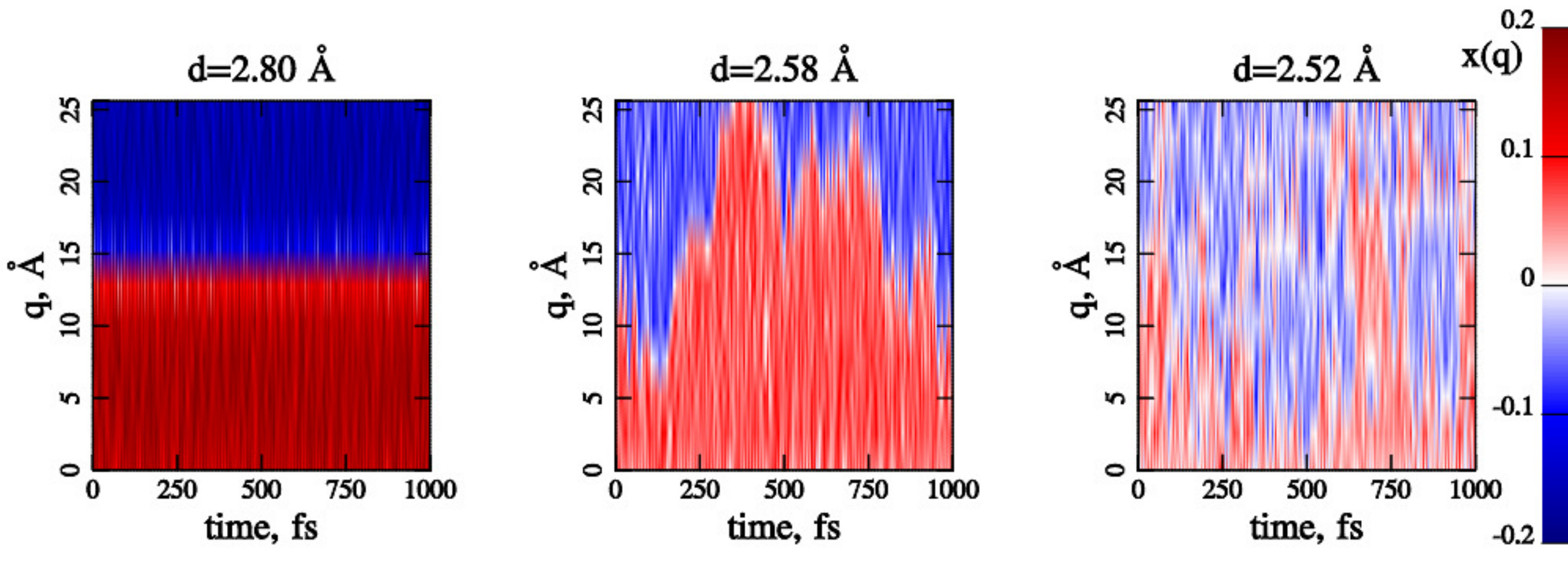}
 \caption{The dynamics of the system. A color referred to the position of the proton. Red means that it is closer to the left oxygen, blue~-- to the right. Leftmost picture~-- stable regime, central~-- random walks, rightmost~-- free hopping. The main parameters of the runs are given at the main text the coupling constant $k_H=0.1$. All parameters except distance $d$ are the same for all runs.\label{dynamics}}
\end{figure}

In order to distinguish between different regimes, we introduce two types of diagnostics. First, we have a simple analysis of the charge dynamics in the system. For each proton we introduce the parameter
\begin{equation}
x_i = q_i - \frac{\phi_i + \phi_{i-1}}{2} .
\end{equation}
Thus, if $x_i < 0$ the $i$-th proton is in the leftmost part of the potential function $V$. For $x_i \geq 0$ the opposite holds. Initially there are 6 protons with $x_i > 0$ and 6 protons with $x_i < 0$. One could also introduce a continuous function $x(q)$ as an interpolation between the discrete values $x_i(q_i)$, such that the point for which $x(q) = 0$ approximately depicts the position of the charge along the line of transport. In Fig.~1 the resulting dynamics of the charge obtained through this method is shown. All the different regimes are easily identified. However, this picture does not allow for an analysis of the average over several runs, and is thus hard to use for investigating the dependence on parameters of the transport process.

\begin{figure}\label{fourier}
 \includegraphics[width=\columnwidth]{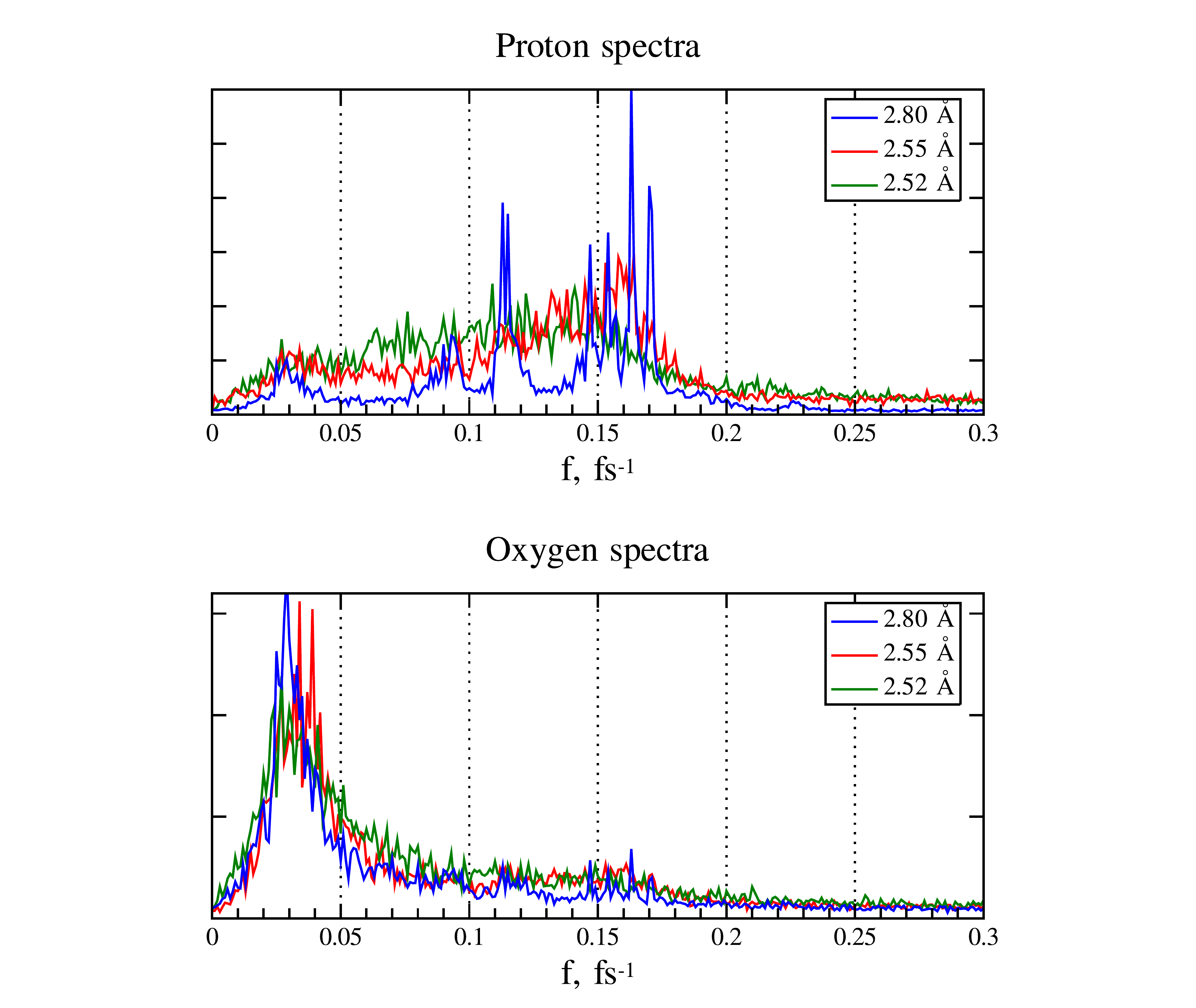}
 \caption{The fourier spectra of the protons and oxygens averaged over all the particles for different initial distances $d$ between oxygens. $d=2.80~\mathrm{\AA}$ (blue lines) refers to the stable regime. Several peaks are observed: low-frequency peak at 0.03~fs$^{-1}$ is due to oxygen motion; high-frequency peaks at 0.095~fs$^{-1}$ and 0.115~fs$^{-1}$ are due to proton oscillations in the wells of potentials $V_1$ (and $V_{N+1}$) and $V$ correspondingly; and finally, high-frequency peaks between 0.14~fs$^{-1}$ and 0.17~fs$^{-1}$ are normal modes of the collective proton oscillations due to interaction between them. $d=2.55~\mathrm{\AA}$ (red lines) refers to the random-walk regime, and the lines are broadened due to the inherent nonlinearity of the system. $d=2.52~\mathrm{\AA}$ (green lines) refers to the free hopping scenario. Here the spectra are even more broad and dominated by intermediate frequencies, determined by the characteristic hopping time.} 
\end{figure}

In order to analyze the averaged behavior we use another technique, based on the fourier analysis of the particle dynamics. In the stable regime, where particles weakly oscillate around equilibrium positions, the spectrum of these oscillations has well defined peaks determined by the frequency of the oscillations. As the distance $d$ decreases the oscillations become more intense and thus the nonlinearity of the interactions come into play, thus giving the spectrum a broadening. This broadening becomes even more pronounced when hopping takes place. Additionally, new spectral lines can appear connected to the characteristic proton hopping frequency between the wells. These lines are dominating in the free-hopping regime. Therefore, the spectrum shows significant changes as the parameters are changed. These changes are easily recognized at Fig.~2.

\begin{figure}\label{dependence_on_distance}
 \includegraphics[width=0.8\columnwidth]{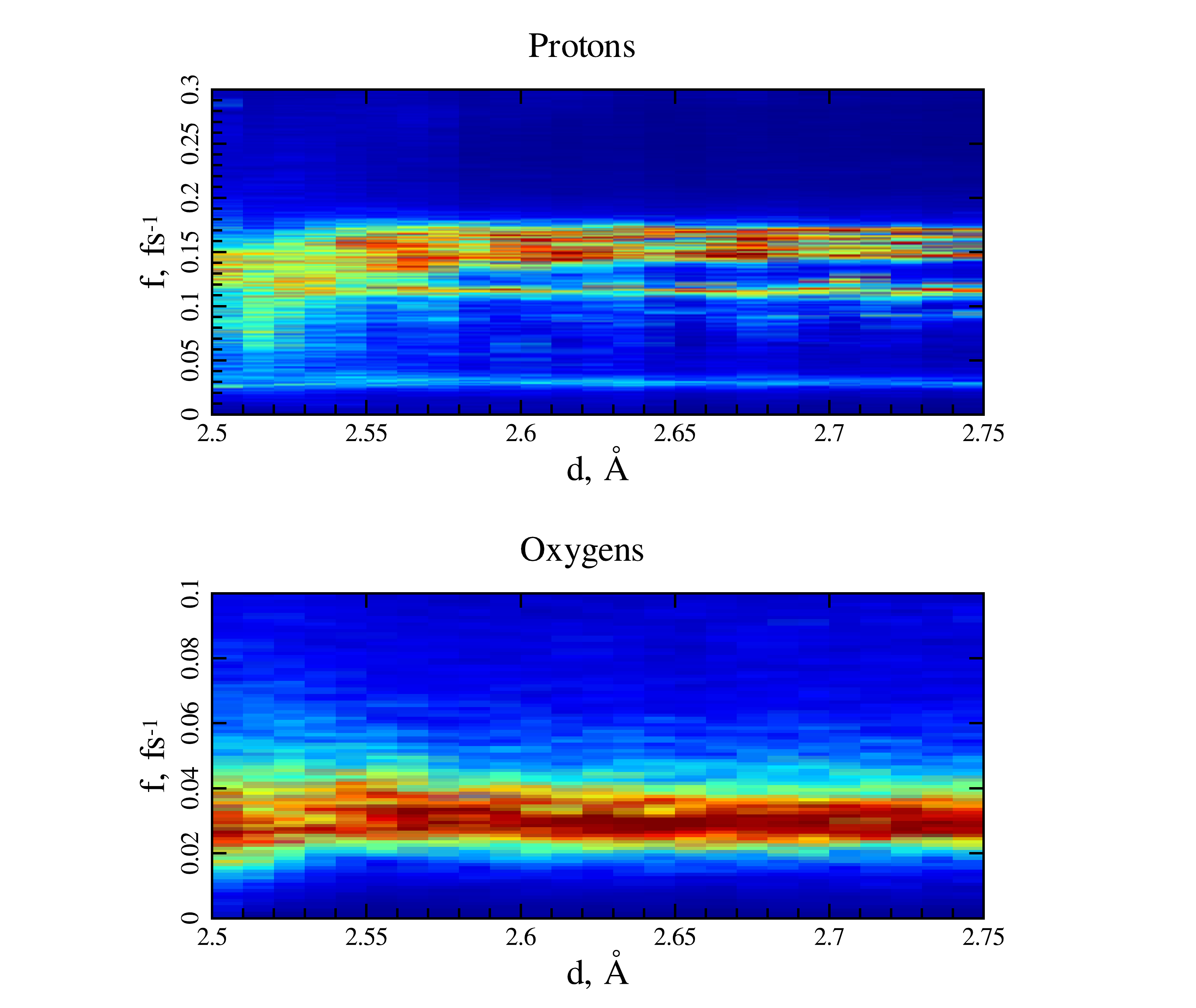}
 \caption{The fourier spectra of the protons (upper panel) and oxygens (lower panel) motion averaged over all the particles and all runs as a function of the distance between the oxygens. The parameters are the same as for Fig.~1} 
\end{figure}

Based on the above technique the dependence on the parameters can be analyzed. In Fig.~3 the dependence of the fourier spectra on the initial distance between the oxygens is shown. As the distance decreases there is a cleat transition from the stable regime to the free-hopping regime through random-walks.

As we can see there is no soliton regime if temperature is dominating over the coupling parameter. This is of course an expected result. Let us consider now the opposite case for which
\begin{equation}\label{soliton condition}
T \ll \frac{k_H B(d-d_0)}{2}
\end{equation}
For the fixed temperature this can be achieved by increasing of coupling constant $k_H$. The dynamics of the system for low and high $k_H$ are shown at Fig.~4. As expected, for strong coupling the random walk regime transforms into a soliton regime in which the proton transport is much faster. It is accompanied by broadening of the proton spectra due to nonlinear nature of the solitons.

\begin{figure}\label{dependence_on_kH}
 \includegraphics[width=0.8\columnwidth]{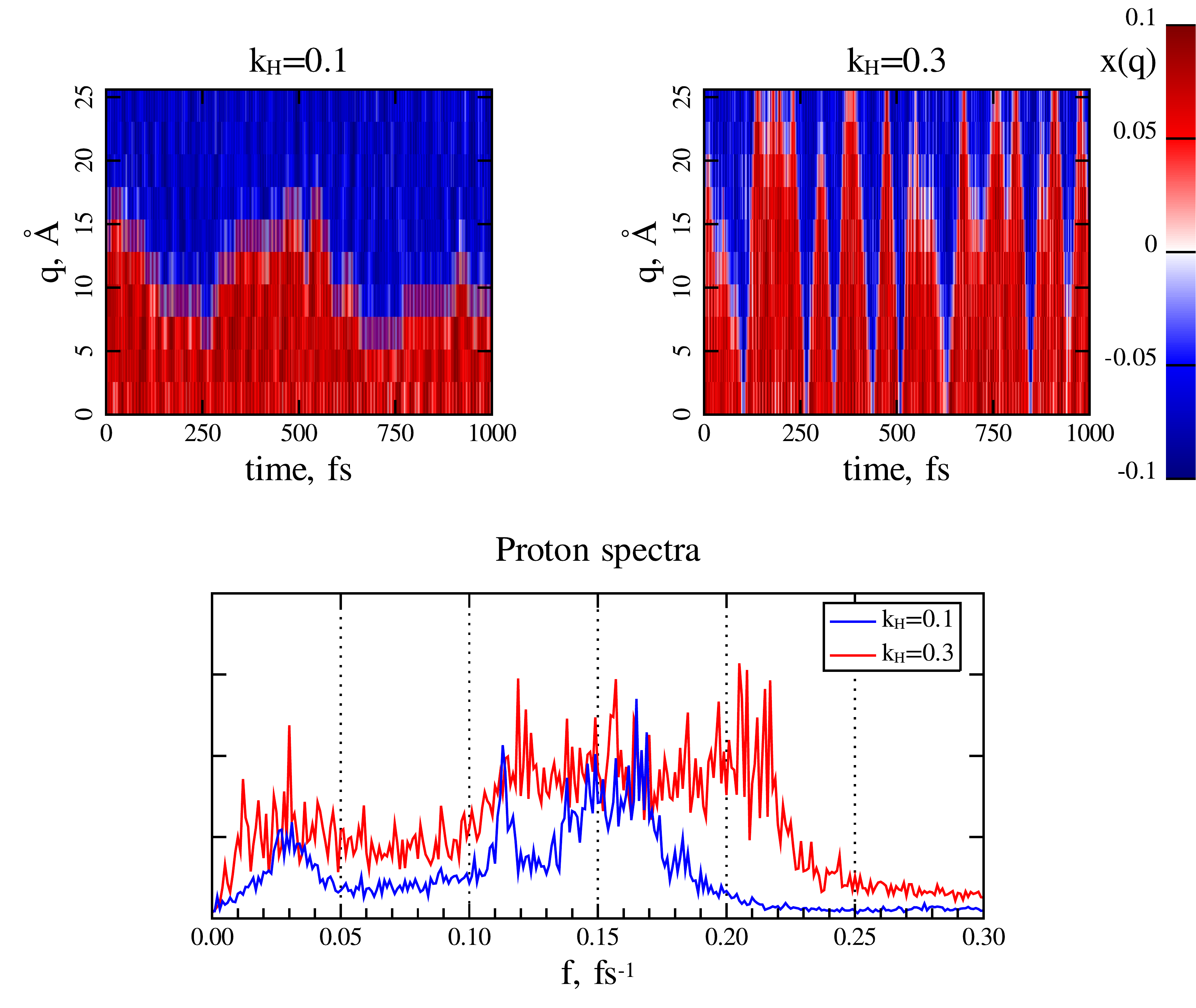}
 \caption{The upper panel: the same as at Fig.~1 but for different $k_H$. The distance between oxygens $d=2.56~\mathrm{\AA}$ is the same for both runs. The lower panel: the fourier spectra of the protons motion averaged over all the protons for the same runs.} 
\end{figure}

\section{Conclusions}
In the current work we have analyzed the different regimes of proton transport in water chains. The analysis is based on a simplified one-dimensional model. The proposed model reflects some of the important and basic properties of the transport process. In particular, it allows for distinguishing four different regimes of interaction: stable, random-walks, free-hopping and solitons. These regimes all have distinct characteristics concerning the transport of protons; in particular, the speed of the transport process is strongly affected by the mode of transport. The model also provides important information on how the transition from one regime to another depends on the system parameters.
If the temperature is low and the coupling between protons is weak, the system is stable and no transport occurs. If the proton energy is comparable to the potential barrier between neighboring molecules two limiting cases can be observed. The first case is realized if the temperature dominates over the coupling. In this case the proton can eventually hop from a hydronium to a neighboring water molecule, and thus the charge transport can be treated as a random-walks. In the opposite case of dominating coupling a soliton is formed in the system and the transport is due to the ballistic propagation of such a proton soliton along the transport direction. And the fourth regime, i.e. free hopping, is observed in the case of low potential barrier and high temperature (as compared to the coupling strength). Protons can freely move between the neighboring molecules by instantaneous hopping from one to another. This latter regime is very unlikely to be realized in any real system at room temperature.

Which of the regimes that occurs in a real system of course depends strongly on the detailed parameters of that particular system. The model used in this work can easily be made more complex and be adapted to different systems. For example, the oscillations can be considered to be three-dimensional and all the potentials involved can be evaluated using e.g. DFT techniques. This will bring the model closer to the reality allowing for more accurate analysis of the particular system of the interest. Such an analysis is left for future research.

\end{document}